\documentclass[twocolumn]{aastex701}
\usepackage{graphicx}
\usepackage{subcaption}
\usepackage{CJK}
\usepackage{amsmath}

\begin{document}
\begin{CJK*}{UTF8}{gbsn}
\title{Singles in Disguise: \textit{Kepler}'s Low-SNR Single-Transiting Systems as Incompletely Detected Multis}
\shorttitle{Singles in Disguise}
\shortauthors{Yahalomi et al.}

\correspondingauthor{Daniel A. Yahalomi}
\email{yahalomi@mit.edu}

\author[0000-0003-4755-584X]{Daniel A. Yahalomi}
\altaffiliation{Juan Carlos Torres Postdoctoral Fellow}
\affiliation{Kavli Institute for Astrophysics and Space Research, Massachusetts Institute of Technology, Cambridge, MA 02139, USA}
\email{yahalomi@mit.edu}

\author[0000-0002-4547-4301]{David Nesvorn\'y}
\affiliation{Southwest Research Institute, 1050 Walnut St, Suite 300, Boulder, CO 80302, USA}
\email{davidn@boulder.swri.edu}

\author[0000-0002-4365-7366]{David Kipping} 
\affiliation{Department of Astronomy, Columbia University, 550 W 120th St., New York NY 10027, USA}
\email{dmk2184@columbia.edu}

\author[0000-0002-9544-0118]{Ben Cassese}
\affiliation{Minor Planet Center, Smithsonian Astrophysical Observatory, 60 Garden Street, Cambridge, MA 02138, USA}
\email{benjamin.cassese@cfa.harvard.edu}

\author[0000-0003-3130-2282]{Sarah C. Millholland} 
\affiliation{Kavli Institute for Astrophysics and Space Research, Massachusetts Institute of Technology, Cambridge, MA 02139, USA}
\affiliation{Department of Physics, Massachusetts Institute of Technology, Cambridge, MA 02139, USA}
\email{sarah.millholland@mit.edu}

\begin{abstract}
NASA's \textit{Kepler} mission found more systems with one transiting planet (``singles'') than expected from those with several (``multis''). This excess was first taken as evidence for a \textit{Kepler} dichotomy -- a second population of intrinsically single or highly inclined systems -- but later work argued that selection effects acting on a single population can reproduce it. Singles also show a puzzling excess of low signal-to-noise ratio (SNR) detections relative to multis. The question arises whether this excess reflects false positives, a selection effect, or an inherent population difference. Here we show, using the \texttt{ExoMOD} population synthesis model and the \textit{Kepler} DR25 catalog, that multis drawn from the same parent population as singles have shorter mean orbital periods, and thus higher-SNR detections on average. Detecting several planets in one system favors compact architectures, so long-period, low-SNR detections are preferentially cataloged as singles, most of them intrinsically multi-planet systems whose other planets went undetected. The low-SNR excess of singles therefore requires neither false positives nor a \textit{Kepler} dichotomy but rather only the geometry and period sensitivity of transit detection.

\end{abstract}

\keywords{}


\section{Introduction}

The \textit{Kepler} mission enabled the first large homogenously studied sample of exoplanetary systems, providing an ideal dataset for statistical characterization of exoplanet system architectures. \textit{Kepler} revealed a rich diversity in planetary multiplicities and orbital configurations. One of the earliest demographic results from \textit{Kepler} was the identification of a discrepancy between the observed number of single-transiting systems and the expectation from multi-planet systems. Analyses of the statistics of \textit{Kepler} multiple-planet systems \citep[e.g.,][]{Lissauer2011, Fang2012, Hansen2013, Ballard2016} originally suggested that the observed distribution of transit multiplicities is difficult to reproduce with a single population of planetary systems having low mutual inclinations. Instead, these studies suggested the need for a two-component model: one population consisting of dynamically cold, high-multiplicity systems, and another consisting of either intrinsically low-multiplicity systems or systems with large mutual inclinations that suppress the probability of multiple transiting planets. This apparent division has become known as the \textit{Kepler dichotomy}.

Subsequent work explored a variety of physical explanations for this dichotomy, including differences in formation pathways and dynamical evolution \citep[e.g.,][]{Johansen2012, Moriarty2016, Hansen2017}. However, these interpretations often require substantial inclination excitation or dynamical instability, which can be difficult to reconcile with long-term system stability. More recent studies have instead argued that the dichotomy may not require two intrinsically distinct populations but rather that observational biases -- especially those related to detection completeness and geometric transit probabilities -- can produce the observed excess of single-transiting systems from an underlying population of multi-planet systems \citep[e.g.,][]{Zhu2018, Sandford2019, Millholland2021, He2020, Gilbert2025}. In this picture, systems identified as ``singles'' may frequently host additional planets that are either not detected or do not transit, blurring the distinction between singles and multis.

Recently, \citet{Lissauer2024} reported a significant difference in the distribution of transit signal-to-noise ratio (SNR) between singles and multis in the \textit{Kepler} catalog. While \textit{Kepler} detected a nearly identical number of planets in single- and multi-transiting systems across an SNR range of 25--180, single-transiting systems exhibit an excess of both low-SNR ($\lesssim$25) and high-SNR ($\gtrsim$180) detections compared to multi-transiting systems. The high-SNR excess is largely driven (and thus can be explained) by hot Jupiters, which are preferentially found in single-transiting systems \citep{Latham2011, Steffen2012}, while the low-SNR excess is less well understood.

As discussed in \citet{Lissauer2024}, several contributing factors may play a role: (1) a larger fraction of singles may be false alarms, particularly in more automated catalogs compared to earlier multi-focused vetting efforts \citep{Rowe2014}; (2) low-SNR candidates cannot be vetted as rigorously, increasing contamination from eclipsing binaries and other false positives, which are more common among targets with only one detected signal \citep{Lissauer2011, Latham2011, Lissauer2012, Lissauer2014, Rowe2014}; (3) in multi-planet systems, detection of additional planets beyond the highest-SNR signal is reduced by pipeline incompleteness, especially when the dominant signal itself has low SNR \citep{Zink2019}; (4) transit-timing variations, which are more frequent in multis, can reduce the pipeline multiple-event statistic and affect detectability; and (5) intrinsic population differences may contribute, as longer-period planets have lower SNR at fixed properties and are geometrically less likely to transit, while planets in multis tend to be clustered in period. 

We can group these possible contributing factors into three categories:
\begin{enumerate}
    \item differences in the false positive (FP) rates of single- and multi-transiting planets (factors 1 and 2);
    \item differences in the true positive (TP) rates --  i.e., the rates at which true planetary signals are recovered for single- and multi-transiting planets (factors 3 and 4);
    \item differences between the intrinsic single- and multi-planet populations -- i.e., a manifestation of and evidence for the \textit{Kepler} dichotomy (factor 5).
\end{enumerate}

Using only \textit{Kepler} data, it is difficult to assess the FP rates, TP rates, and intrinsic populations, as one only has access to the observed transiting planets, not the underlying planetary population. Several forward models of the intrinsic planetary population have been developed, ranging from semi-analytic architecture models \citep{Zhu2018} to full population synthesis frameworks such as EPOS \citep{Mulders2018} and SysSim \citep{He2019, He2020, He2021}. These models constrain the underlying population by comparing simulated observations against the \textit{Kepler} catalog; notably, both \citet{Zhu2018} and \citet{He2020} found that the observed multiplicity distribution can be reproduced by systems whose mutual inclinations anti-correlate with intrinsic multiplicity to reproduce the observed number of single-transiting systems without invoking a separate population of intrinsic singles. We build on these results by asking specifically whether the \emph{SNR structure} of the single--multi difference reported by \citet{Lissauer2024} -- the low-SNR excess among singles -- likewise arises from selection rather than from a distinct single-planet population.

We use the new population synthesis model \texttt{ExoMOD}, which is explained in depth in \citet{ExoMOD1, ExoMOD2}. In short, \texttt{ExoMOD} defines analytic representations of the distributions that describe the intrinsic properties of \textit{Kepler}'s close-in planetary systems. These include prescriptions for their multiplicity, planetary radii, orbital periods and eccentricities, and mutual inclinations assigned via an Angular Momentum Deficit (AMD) stability framework \citep{Laskar2017,Petit2017, He2020}. \texttt{ExoMOD} then generates statistical samples of planetary systems around the observed catalog of \textit{Kepler} FGK dwarf targets. Model planets are subjected to both the geometric transit condition and an SNR-based detection efficiency model \citep{Fressin2013, KippingSandford2016}, producing sets of model-detected planets. The model parameters are then optimized, and competing intrinsic models compared, with \texttt{MultiNest} \citep{Feroz2008, Feroz2009} by fitting the observed multiplicity, orbital period, and period ratio distributions of the \textit{Kepler} DR25 catalog, restricted to FGK dwarf hosts ($0.5 < R_{\rm p}/R_\oplus < 7$, and $3 < P_{\rm orb} < 300$ days). 

As shown in \citet{ExoMOD1}, \texttt{ExoMOD}'s intrinsic model selected via Bayesian evidence reproduces the observed \textit{Kepler} multiplicity, orbital-period, period-ratio, and planetary-radius distributions, as well as higher-order architecture metrics including radius partitioning, radius monotonicity, and gap complexity (the irregularity of orbital spacing). Critically for this work, \texttt{ExoMOD} draws the innermost orbital period from the same distribution irrespective of system multiplicity -- it contains no multiplicity-dependent period law -- yet still reproduces the statistically distinct period distributions of observed singles and multis, which \citet{ExoMOD1} attribute to the finite \textit{Kepler} baseline. Throughout this paper, we adopt the preferred \texttt{ExoMOD} model of \citet{ExoMOD1} (their auxiliary model $\mathcal{M}_{228}$: Poisson intrinsic multiplicity, $f_{\rm AMD} = 0.62$, transition to correlated period ratios at $m_* = 4.5$), which provides the best fit when constrained by gap complexity -- drawing on the ten publicly available $\mathcal{M}_{228}$ realizations generated with different random seeds.\footnote{\url{https://www.boulder.swri.edu/~davidn/ExoMOD/}}

While both frameworks employ forward modeling to constrain intrinsic architectures, and many of the choices made in \texttt{ExoMOD} were inspired by those made in \texttt{SysSim}, there are some subtle differences of choices in their implementations. First, \texttt{ExoMOD} explores multiple multiplicity distribution models (both Poisson and Zipfian); the observed multiplicity and higher-order architecture metrics are robustly reproduced across these different underlying distributions, validating the framework's physical assumptions.  Second, following the peas-in-a-pod correlation observed in \textit{Kepler} systems, \texttt{ExoMOD} draws outer planet radii from a Student-\textit{t} distribution centered on the inner planet, whereas \texttt{SysSim} employs log-normal distributions; the Student-\textit{t} approach allows for heavier tails that replicate observed radius ratios. Third, \texttt{ExoMOD} places the innermost planet via a power-law orbital period distribution, while outer planets are positioned at Hill-radius-scaled multiples drawn from a normal distribution. In contrast, \texttt{SysSim} distributes all planets using log-normal orbital period distributions; \texttt{ExoMOD}'s hybrid approach effectively reproduces both the uniformity and gap complexity characteristics of \textit{Kepler} systems. \texttt{ExoMOD}'s spacing scheme naturally produces the observed correlation wherein smaller planets exhibit tighter orbital period separations (lower $P_j/P_{j-1}$ ratios) than larger planets. This physically motivated methodology captures the interplay between planet masses, orbital stability, and observed architecture. Fourth, \texttt{ExoMOD} employs a 2D-grid binning likelihood that explicitly preserves correlations between multiplicity and other parameters (e.g., orbital period), whereas \texttt{SysSim} marginalizes over independent 1D distributions. Fifth, while both frameworks adopt AMD stability criteria, \texttt{ExoMOD} treats the AMD fraction $f_{\rm AMD}$ as a free parameter that can be constrained by higher-order architecture metrics like gap complexity, recovering intermediate AMD values ($f_{\rm AMD} \sim 0.6$) that reflect realistic formation pathways involving resonant chains and subsequent instabilities, whereas \texttt{SysSim} assumes systems exist at the critical AMD value ($f_{\rm AMD} = 1$).

In Section~\ref{sec: SNR_equations} we present the mathematical formalism of signal-to-noise in transit surveys. In Section~\ref{sec: catalog_building} we describe the construction of the catalog of single- and multi-transiting systems that we use throughout the analysis that follows. Section~\ref{sec: FP} tests whether false positives can explain the low-SNR excess, using both the DR25 catalog and the false positive-free \texttt{ExoMOD} forward model. Section~\ref{sec:kepler_population} examines the observed differences between singles and multis directly from the \textit{Kepler} data, isolating orbital period as the primary driver of the SNR offset via a Monte Carlo resampling experiment and nearest-neighbor matched-filtering. Section~\ref{sec: intrinsic} uses \texttt{ExoMOD} to ask whether the observed single--multi period difference reflects an intrinsic population-level difference or is induced by detection. Section~\ref{sec: discussion} interprets these results in terms of the period dependence of transit detectability, and Section~\ref{sec: conclusions} 
summarizes our findings.

\section{Background}

\subsection{Theoretical Transit SNR} \label{sec: SNR_equations}
Transit detectability depends on the number of observed transits and the per-transit signal-to-noise ratio. For a fixed observing baseline $T_{\rm obs}$ (as for \textit{Kepler}), the number of transits scales as
\begin{equation}
N_{\rm tr} \approx \frac{T_{\rm obs}}{P},
\end{equation}
such that coherent folding increases detection significance as
\begin{equation}
{\rm SNR} \propto \sqrt{N_{\rm tr}} \propto \frac{1}{\sqrt{P}}.
\end{equation}

If we define the per-transit SNR as ${\rm SNR}_{\rm tr}$, we then have

\begin{equation} \label{eq:snr_tot}
{\rm SNR} \propto \frac{{\rm SNR}_{\rm tr}}{\sqrt{P}}.
\end{equation}

\citet{Kipping2023} derives the optimal matched-filter signal-to-noise ratio for a trapezoidal transit shape, per transit, as
\begin{equation}
{\rm SNR}_{\rm tr}
\;=\;
\frac{\delta}{\sigma_0}
\sqrt{\frac{T_{14}+2T_{23}}{3}},
\end{equation}
where $\delta$ is the transit depth, given in the small-planet limit by
\begin{equation}
\delta \;\approx\; \left(\frac{R_p}{R_\star}\right)^2,
\end{equation}
$\sigma_0$ is the photometric noise per unit time, $T_{14}$ is the total transit duration from first to fourth contact, and $T_{23}$ is the full-transit duration from second to third contact.

Defining an effective duration,
\begin{equation}
T_{\rm box} \;\equiv\; \frac{T_{14}+2T_{23}}{3},
\end{equation}
this can be written compactly as
\begin{equation}
{\rm SNR}_{\rm tr}
\;=\;
\frac{\delta}{\sigma_0}\,\sqrt{T_{\rm box}}.
\end{equation}
In this sense, $T_{\rm box}$ represents an SNR-equivalent duration: the duration of a box-shaped transit that yields the same optimal matched-filter SNR as the trapezoidal signal.

To relate this quantity to orbital period, we assume (i) circular orbits, (ii) a fixed stellar mean density, (iii) the small-planet limit ($R_p \ll R_\star$), and (iv) a representative, period-independent distribution of impact parameters. Under these assumptions, the total transit duration scales as
\begin{equation}
T_{14} \;\propto\; \frac{R_\star}{a} P,
\end{equation}
where $a$ is the semi-major axis. Using \textit{Kepler}'s third law ($a \propto P^{2/3}$ for fixed stellar density), this gives
\begin{equation}
T_{14} \;\propto\; P^{1/3}.
\end{equation}

Since $T_{\rm box}$ differs from $T_{14}$ only by a shape-dependent factor of order unity (set by the ratio $T_{23}/T_{14}$, which depends primarily on impact parameter), it follows that, to leading order,
\begin{equation}
T_{\rm box} \;\propto\; P^{1/3}.
\end{equation}
Consequently, the per-transit SNR scales as
\begin{equation} \label{eq:snr_transit}
{\rm SNR}_{\rm tr} \;\propto\; \frac{\delta}{\sigma_0} \, P^{1/6}.
\end{equation}

Combining Equation~\ref{eq:snr_tot} with Equation~\ref{eq:snr_transit}, we get our final SNR
\begin{equation} \label{eq:snr_final}
{\rm SNR}
\;\propto\;
\frac{\delta}{\sigma_0}\,P^{-1/3}.
\end{equation}

This scaling is consistent with classical predictions of transit survey yields.  \citet{BeattyGaudi2008} derived equivalent period dependence in their predictions for photometric survey yields through $\chi^2 = N_{\rm tr}(\delta/\sigma)^2$; since ${\rm SNR} \propto \sqrt{\chi^2}$, the number of observable transits  scales inversely with period while individual transit SNR scales as $P^{1/6}$,  yielding the same $P^{-1/3}$ dependence.

This scaling highlights the three primary physical factors that determine transit detectability and SNR for a given observing baseline and instrument. First, the SNR scales with planet size through the transit depth, ${\rm SNR} \propto \delta \propto (R_p/R_\star)^2$, making larger planets significantly easier to detect. Second, the SNR decreases with orbital period as ${\rm SNR} \propto P^{-1/3}$, reflecting the combined effects of fewer observed transits and only weak gains in per-transit SNR at longer durations. Finally, the SNR is inversely proportional to the stellar photometric noise, ${\rm SNR} \propto \sigma_0^{-1}$, such that quieter stars (or improved instrumental precision) directly enhance detectability. 

These three dependencies -- planet radius, orbital period, and stellar noise -- represent the primary physical factors that can differ between planetary systems and thus are the three values that should be investigated in studying the SNR differences between planetary populations observed by the same telescope and instrument.

\subsection{Catalog Extraction and Classification of Singles vs.\ Multis} \label{sec: catalog_building}

We construct our \textit{Kepler} sample by combining the DR25 KOI table \citep{koidr25}\footnote{Accessed on 2026-03-15 at 09:08, returning 8203 rows.} (includes the full Q1--Q17) downloaded from the NASA Exoplanet Archive \citep{Christiansen2025} with the DR25 stellar catalog \citep{Mathur2017}.\footnote{We use the DR25 stellar properties \citep{Mathur2017} rather than the more precise Gaia-based values of \citet{Berger2020}. None of the quantities we analyze depends on the adopted stellar radius: the MES, $P$, and CDPP$_{6\,\mathrm{hr}}$ are \textit{Kepler} pipeline products, and $R_p/R_\star$ (\texttt{koi\_ror}) is fit directly to the light curve. Stellar properties enter only through the $\log g$, $R_\star$, and $R_p$ cuts that define the sample, and adopting DR25 keeps our sample consistent with the stellar properties used by \texttt{ExoMOD} \citep{ExoMOD1}.} We perform an inner join on \texttt{kepid} so that each retained planet candidate is matched to a host star with available stellar parameters.

We define the working \textit{Kepler} dataset by applying a sequence of cuts to this merged table. First, we require \texttt{koi\_disposition != FALSE POSITIVE}, i.e., that the ``Exoplanet Archive Disposition'' does not identify the planet as a false positive. We then restrict the host stars to a FGK dwarf-star sample by requiring $\log g > 4.0$ (cgs) and $0.7 < R_\star/R_\odot < 1.6$ where $R_\odot$ is the Solar radius. For the planetary candidates themselves, we require $0.5 < R_p/R_\oplus < 7.0$, where $R_\oplus$ is the Earth radius, and $3~{\rm d} < P_{\rm orb} < 500~{\rm d}$. This catalog selection largely matches that of \citet{ExoMOD1}, with two key differences. First, we exclude KOIs for which the ``Exoplanet Archive Disposition'' is reported as \texttt{FALSE POSITIVE}, whereas \citet{ExoMOD1} require that the ``Disposition using Kepler Data'' be \texttt{CANDIDATE}.\footnote{``Disposition using Kepler Data'' reflects the \textit{Kepler} data alone following the false positive tests as described in \citet{Batalha2013}, whereas ``Exoplanet Archive Disposition'' aggregates all available evidence, including follow-up observations. Our cut is therefore the more conservative of the two: we discard any KOI identified as a false positive by any dataset.} Second, we adopt an upper orbital period limit of 500 days, compared to the 300-day limit used in \citet{ExoMOD1}. The extended 500-day limit is motivated by a balance between retaining long-period planets of interest and avoiding excessive contamination by false positives. Periods beyond $\sim$500 days are known to be highly contaminated because such signals typically only marginally satisfy the \textit{Kepler} pipeline requirement of three observed transits \citep{Mullally2015, Zink2019}. This choice therefore allows us to probe longer-period planets while remaining within a regime where the candidate sample is still reliable.

In addition, we remove candidates with orbital periods near 372 days to mitigate contamination associated with the \textit{Kepler} spacecraft's orbital period. Specifically, we exclude planets with periods within $\pm 25$ days of 372 days, where spurious signals are known to arise due to systematic effects tied to the spacecraft's yearly observing geometry \citep[e.g.,][]{Mullally2018, Thompson2018}. This step reduces the impact of known instrumental false positives that cluster near the \textit{Kepler} orbital period.

We define the observed multiplicity of a system using the host-star identifier \texttt{kepid}. After all sample cuts are applied, we group the retained candidates by \texttt{kepid} and assign each candidate a multiplicity equal to the number of surviving candidates associated with that star. Systems with observed multiplicity equal to one are classified as \textit{singles}, while those with observed multiplicity of two or more are classified as \textit{multis}. Because observed multiplicity is computed only after the full sample selection, excluded candidates do not contribute to the system counts. This yields a total of $n(\mathcal{D}_{\rm kep}) = 2{,}557$ planets. Let $\mathcal{D}_{\rm kep, m}$ denote the subset of planets in systems of observed multiplicity $m$. We find $n(\mathcal{D}_{\rm kep,1}) = 1{,}485$ singles, $n(\mathcal{D}_{\rm kep,2}) = 604$, $n(\mathcal{D}_{\rm kep,3}) = 300$, $n(\mathcal{D}_{\rm kep,4}) = 116$, $n(\mathcal{D}_{\rm kep,5}) = 40$, and $n(\mathcal{D}_{\rm kep,6}) = 12$, with no systems of multiplicity $m \geq 7$.

Combining all systems with $m \geq 2$, this corresponds to 1,485 planets in single-planet systems and 1,072 planets in multi-planet systems. We emphasize that multiplicity here refers strictly to the number of \textit{Kepler}-detected transiting planet candidates within each system. Thus, ``single'' and ``multi'' denote single- and multi-transiting systems, respectively, and do not imply that these systems host only these planets intrinsically. Additional non-transiting planets may be present but are not included in this classification.

Throughout this work we adopt the Multiple Event Statistic (MES; \texttt{koi\_max\_mult\_ev} in the DR25 KOI table) as our proxy for the transit signal-to-noise ratio, and all references to SNR (when discussing the $\mathcal{D}_{\rm kep}$ population) hereafter refer to this quantity. The MES is the detection statistic of the \textit{Kepler} pipeline's wavelet-based matched filter, constructed by combining the individual transit events of a candidate periodic signal such that, for well-behaved noise, its expected value equals the SNR of the full transit sequence \citep{Jenkins2002}. The pipeline defined a threshold-crossing event as a periodic signal with ${\rm MES} \geq 7.1$, a threshold chosen to keep the expected number of purely statistical false alarms across the mission small \citep{Jenkins2002}. This expectation, however, was derived under the assumption of white noise; \textit{Kepler} light curves contain significant correlated noise, and injection-recovery and reliability studies have shown that false alarms remain common in the range $7.1 \leq {\rm MES} \lesssim 10$ \citep{Thompson2018, Hsu2019}. In addition, candidates near the detection threshold lack the signal needed for precise centroid analysis and other vetting diagnostics that separate transiting planets from astrophysical false positives such as eclipsing binaries, so more stringent MES cuts are required to construct high-purity candidate samples \citep{Lissauer2024}. We prefer the MES over the transit-model signal-to-noise ratio from the data-validation fits (\texttt{koi\_model\_snr}) because the detection threshold and the DR25 injection-recovery completeness measurements \citep{Christiansen2020} are both defined in terms of the MES; adopting it as our SNR proxy keeps our analysis self-consistent with the pipeline's actual detection criterion.\footnote{We additionally verified that repeating the analysis that follows with the reported transit-model SNR (\texttt{koi\_model\_snr}) yields similar results.}

\section{Can False Positives Explain the Low-SNR Excess?} \label{sec: FP}
 
\begin{figure*}[htb!]
    \centering
    \includegraphics[width=\linewidth]{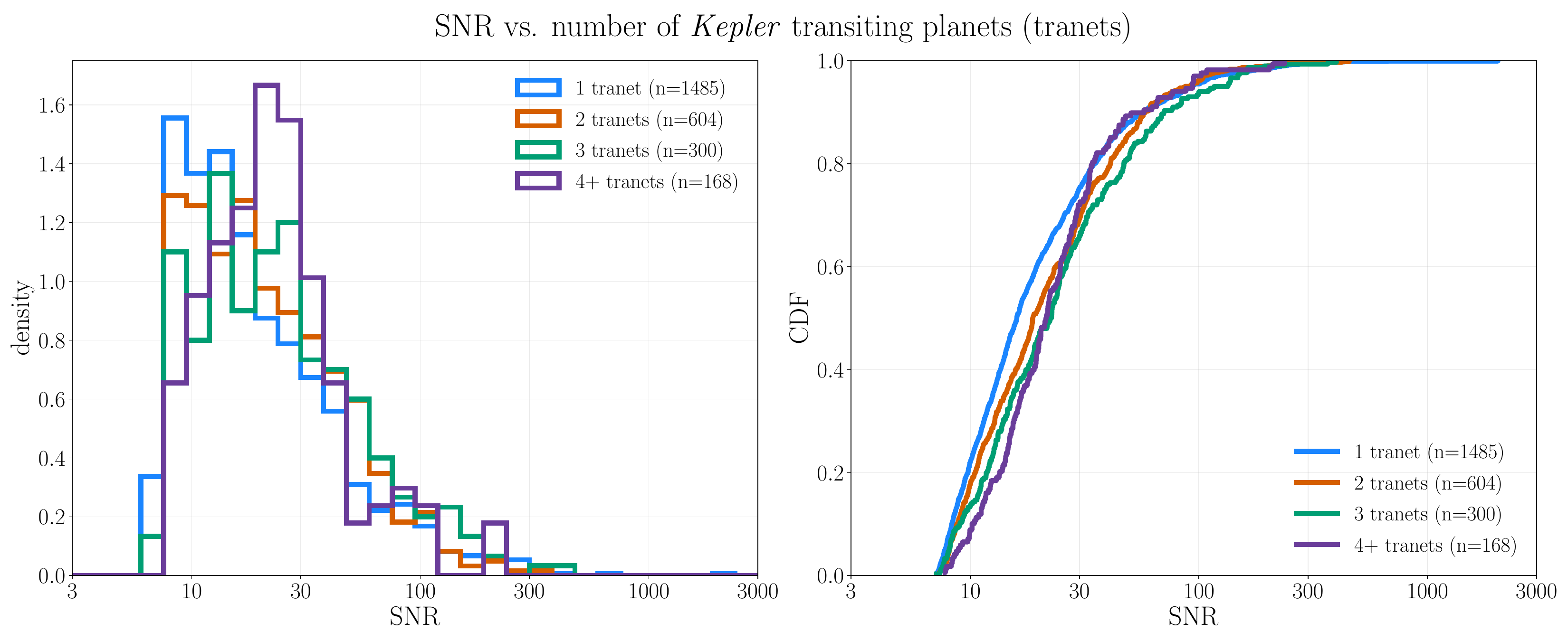}
    \caption{Distributions (left: normalized histograms; right: empirical CDFs) of the SNR of \textit{Kepler} planet candidates, split by the observed transiting multiplicity of their host systems ($m_{\rm obs} = 1$, 2, 3, and $4+$). The fraction of low-SNR detections decreases monotonically with multiplicity: singles are the most skewed toward low SNR, followed by doubles, triples, and $4+$ systems.}
    \label{fig:mult_split_trends}
\end{figure*}
 
\begin{figure*}[htb!]
    \centering
    \includegraphics[width=\linewidth]{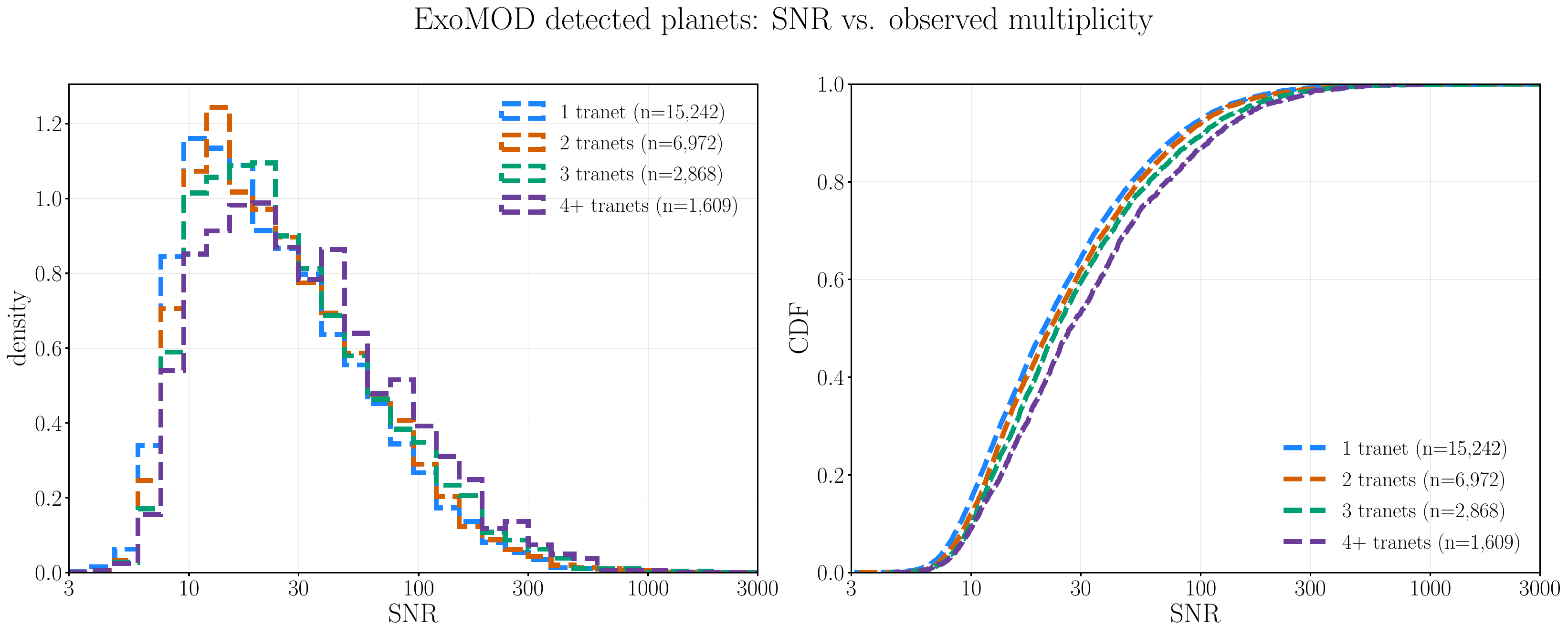}
    \caption{Distributions (left: normalized histograms; right: empirical CDFs) of the SNR of \texttt{ExoMOD} forward-modeled detected planets, split by observed transiting multiplicity ($m_{\rm obs}=1$, 2, 3, and $4+$). As in the \textit{Kepler} data (Figure~\ref{fig:mult_split_trends}), the low-SNR fraction declines monotonically with multiplicity; because \texttt{ExoMOD} contains no false positives by construction, this ordering must arise from selection and population-level effects alone.}
    \label{fig:exomod_snr_split}
\end{figure*}
 
We first test whether the difference between the SNR distributions of singles and multis could be produced by false positives. False positive contamination is expected to be concentrated heavily in singles: eclipsing binaries and instrumental artifacts rarely masquerade as multiple coherent periodic signals on the same target, and the detection of additional transiting planets in a system is itself strong statistical evidence against a false positive \citep{Lissauer2012, Lissauer2014}. This hypothesis therefore makes a testable prediction: the low-SNR excess should be confined to $m = 1$, so that the SNR distributions of $\mathcal{D}_{\rm kep,m}$ for $m \geq 2$ are statistically consistent with one another and $\mathcal{D}_{\rm kep,1}$ stands out as the lone outlier. If instead the SNR distributions continue to shift with increasing $m$, contamination cannot be the sole cause, and some mechanism that varies smoothly with multiplicity must also be at work. We therefore examine the SNR distributions as a function of $m$, rather than only across the single--multi divide.
 
We split the systems in our \textit{Kepler} catalog by observed transiting multiplicity into groups with 1, 2, 3, and 4 or more (4+) transiting planets (tranets), and compare the SNR distributions of the individual tranets in each group (Figure~\ref{fig:mult_split_trends}). To quantify the comparison, we define the low-SNR fraction of each group as the probability that a tranet falls below an SNR threshold $S$ given that it belongs to a system of observed multiplicity $m$,
\begin{equation}
    f_m(S) \equiv P\!\left({\rm SNR} < S \,\middle|\, m_{\rm obs} = m\right) \, ,
    \label{eq:fm}
\end{equation}
and adopt $S = 20$ as our fiducial threshold, writing $f_m \equiv f_m(20)$. The low-SNR fraction declines monotonically with multiplicity: $f_m$ falls from 61\% for singles to 52\%, 47\%, and 46\% for doubles, triples, and 4+ systems, and $f_m(15)$ from 47\% to 38\%, 34\%, and 27\%. At the ${\rm SNR} \lesssim 25$ threshold highlighted by \citet{Lissauer2024}, the singles excess persists ($f_1(25) = 69\%$ versus 58--61\% for the multis), although the ordering among the multiplicity groups breaks down above ${\rm SNR} \approx 22$. Because the number of tranets per group falls steeply with multiplicity, from 1,485 singles to 604, 300, and 168 in the double, triple, and 4+ groups, the differences among the multis are only modestly larger than their counting uncertainties. On its own, the observed sample therefore suggests, but does not firmly establish, that the decline extends beyond the singles.
 
To test whether such a decline is expected in the absence of false positives, we repeat the analysis on the synthetic detected population from the \texttt{ExoMOD} forward model \citep{ExoMOD1}, which by construction contains no false positives: every simulated detection is a true planet, drawn from the fitted population model and passed through a \textit{Kepler}-like selection function (the geometric transit condition and an SNR-dependent detection probability). The simulated population comprises 26,691 detected planets across ten realizations (15,242, 6,972, 2,868, and 1,609 in the four multiplicity groups), roughly ten times the \textit{Kepler} sample. As shown in Figure~\ref{fig:exomod_snr_split}, the \texttt{ExoMOD} detected planets exhibit the same ordered shift toward higher SNR with increasing observed multiplicity: the median SNR rises from 20.8 for singles to 22.1, 23.6, and 27.0 for doubles, triples, and 4+ systems, and the low-SNR fraction $f_m^{\rm ExoMOD}$ falls from 49\% to 46\%, 42\%, and 37\% ($f_m^{\rm ExoMOD}(15)$ from 36\% to 33\%, 29\%, and 25\%). With this larger sample, the ordering holds at every threshold we test, including $S = 25$, consistent with the breakdown near ${\rm SNR} \approx 22$ in the observed sample being a consequence of the small size of the highest-multiplicity groups. Because no false positives are present in the simulation, this trend must arise entirely from selection and population-level effects.
 
False positives are therefore not needed to produce a low-SNR fraction that declines with multiplicity. \texttt{ExoMOD} is not fit to the SNR distribution, and its detection-efficiency model and our DR25 selection differ in detail; its low-SNR fractions are lower than the observed ones for every multiplicity group, so we compare trends rather than absolute values. One difference in the trends remains: the drop from singles to doubles is larger in the data (61\% to 52\%) than in the model (49\% to 46\%). This could reflect residual false positives among the singles, or the 300--500\,d planets in our catalog (Section~\ref{sec: catalog_building}), which \texttt{ExoMOD} does not model and which are preferentially low-SNR singles. Either would add to the singles' low-SNR excess, but neither is required to produce the ordered decline, which appears even in a population with no false positives.

\section{Observed differences between Singles and Multis} \label{sec:kepler_population}

Having found (Section~\ref{sec: FP}) that false positives are not required to produce the low-SNR excess among singles, we now ask whether it instead reflects differences in the observed planetary characteristics of singles and multis. Equation~\ref{eq:snr_final} suggests that SNR depends on three observables -- the radius ratio $R_p/R_\star$ (through the transit depth), the orbital period $P$, and the stellar photometric noise $\sigma_0$ -- so any SNR difference between the two populations that is not due to false positives must arise from a difference in one, or a combination, of these three quantities. The aim of this section is to isolate which of these quantities plays a dominant role.

We first focus on the low-SNR population in which the excess resides, the 1,020 singles and 645 multis in $\mathcal{D}_{\rm kep}$ with SNR $<25$, and compare the median of each observable between the two groups. The three observables behave differently. Singles are longer-period, with a median period of $13.8$~d against $12.1$~d for multis ($14\%$ longer), and they are found around noisier stars, with a median $\mathrm{CDPP}_{\rm 6\,hr}$ of $123$~ppm against $107$~ppm ($15\%$ higher) -- both differences act to lower the SNR of singles. Their radius ratios show the smallest difference of the three, and in the opposite direction: the median $R_p/R_\star$ is $0.0153$ for singles versus $0.0143$ for multis, i.e.\ singles are about $7\%$ \emph{larger}. Because larger planets produce deeper, higher-SNR transits, this radius difference works against the observed offset rather than for it, and so cannot be its source.

Two of the three observables -- orbital period and stellar noise -- therefore differ in the direction that lowers the SNR of singles, while radius is excluded by its sign. Interestingly, for orbital period, the single--multi separation grows toward the low-SNR regime ($14\%$ in the ${\rm SNR}<25$ sub-sample versus $8\%$ in the full sample), tracking the excess. In contrast, the noise difference is comparable in both the low-SNR sub-sample and the full sample($15$--$16\%$ in both cases). A comparison of medians alone, however, cannot disentangle contributions from observables that may themselves be correlated. We therefore isolate the role of orbital period directly, first with a Monte Carlo resampling experiment that holds all non-period contributions fixed and swaps only the period distribution (Section~\ref{sec:mc_snr}), and then, more rigorously, with a matched-filtering analysis that controls simultaneously for $R_p/R_\star$ and stellar noise in addition to period (Section~\ref{sec:snr_match_examples}).

\subsection{Monte Carlo Resampling of SNR Distributions}
\label{sec:mc_snr}

\begin{figure*}[htb!]
    \centering
    \includegraphics[width=\textwidth]{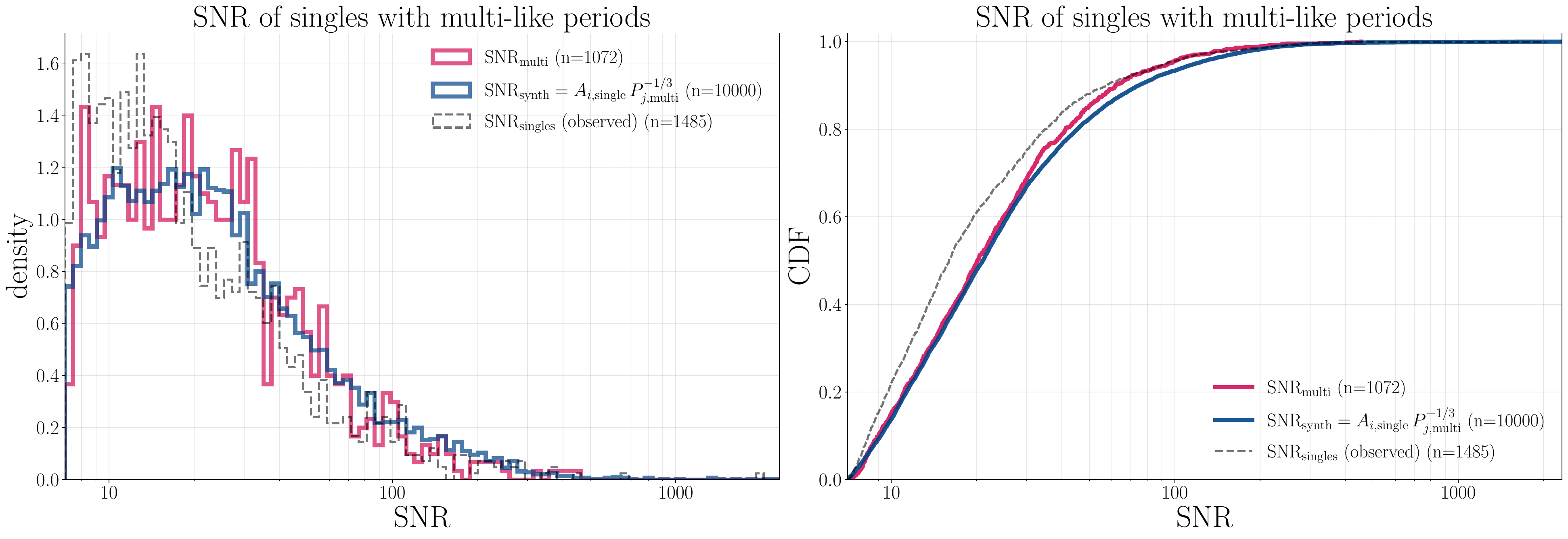}
    \caption{Monte Carlo resampling experiment in which values of $A_{i,{\rm single}}$ (see Equation~\ref{eq:mc_single}) are drawn from the single-transiting population while orbital periods $P_{j,{\rm multi}}$ are drawn from the multi-transiting population. For each Monte Carlo realization, the synthetic SNR is computed as ${\rm SNR}_{\rm synth} = A_{i,{\rm single}}\,P_{j,{\rm multi}}^{-1/3}$. The resulting synthetic SNR distribution closely resembles the observed SNR distribution of multis, suggesting that differences in orbital-period distributions drive the observed SNR offset between singles and multis.}
    \label{fig:mc_swap_example}
\end{figure*}

To investigate the role of orbital period in shaping the observed signal-to-noise ratio (SNR) distributions of single- and multi-transiting systems, we perform a Monte Carlo resampling experiment motivated by the theoretical scaling presented in Equation~\ref{eq:snr_final}. We re-write this equation as
\begin{equation}
{\rm SNR} = A\,P^{-1/3}.
\end{equation}

The quantity $A$ thus encapsulates all non-period-dependent contributions to the transit SNR, including transit depth ($\delta$), stellar photometric noise ($\sigma_0$), and additional effects not explicitly captured by the leading-order period scaling.

For each single-transiting planet candidate $i$, with measured signal-to-noise ratio ${\rm SNR}_{i,{\rm single}}$ and orbital period $P_{i,{\rm single}}$, we compute

\begin{equation} \label{eq:mc_single}
A_{i,{\rm single}}
=
{\rm SNR}_{i,{\rm single}}
\,P_{i,{\rm single}}^{1/3}.
\end{equation}

By construction, $A_{i,{\rm single}}$ isolates everything about a single-transiting candidate except its orbital period. We can therefore generate a Monte Carlo population of hypothetical single-transiting systems placed on multi-like orbits: at each draw we combine two quantities, a value of $A_{i,{\rm single}}$ drawn at random from the single-transiting population and an orbital period $P_{j,{\rm multi}}$ drawn independently from the multi-transiting population. Their combination defines a synthetic SNR for a single planet ($i$) had it been observed with the orbital period of a typical multi-transiting planet ($j$), holding all else fixed:

\begin{equation}
{\rm SNR}_{i,{\rm synth}}
=
A_{i,{\rm single}}
\,P_{j,{\rm multi}}^{-1/3}.
\end{equation}

Repeating this over many $(i,j)$ combinations builds up an entire synthetic population, defined formally as

\begin{equation} \label{eq:snr_synth_pop}
\begin{split}
{\rm SNR}_{\rm synth} = \big\{ A_{i,{\rm single}}^{(k)} \left(P_{j,{\rm multi}}^{(k)}\right)^{-1/3} \\
\mid k = 1, 2, \ldots, N_{\rm draw} \big\},
\end{split}
\end{equation}
\noindent which represents the SNR distribution that single-transiting systems would have shown had they instead been observed with periods drawn from a typical multi-planet orbit. By holding the distribution of $A$ fixed to that of real singles while swapping in the period distribution of real multis, this synthetic population isolates whether the SNR offset -- in particular the excess of low-SNR singles -- is driven by the $P$ term or by the $A$ term that encapsulates everything else.


In the experiment shown in Figure~\ref{fig:mc_swap_example}, we draw $N_{\rm draw} = 10{,}000$ values from the single-transiting population while drawing orbital periods from the multi-transiting population. The resulting synthetic SNR distribution closely resembles the observed SNR distribution of multis. This result suggests that differences in the orbital-period distributions between singles and multis contribute a substantial fraction of the observed SNR offset.

However, it is important to note that this resampling framework implicitly assumes that the non-period-dependent normalization factor $A$ is statistically independent of orbital period within each population, which we do not expect to be true in practice. So this Monte Carlo resampling procedure should not be interpreted as a fully self-consistent generative model of the observed SNR distributions. Rather, it serves as a heuristic diagnostic demonstrating that the orbital-period distribution contains a substantial fraction of the information needed to reproduce the observed differences between singles and multis. To test this conclusion without relying on the independence assumption, we now turn to a matched-filtering analysis (Section~\ref{sec:snr_match_examples}) that controls for period -- along with $R_p/R_\star$ and stellar noise -- by directly matching systems in these observable quantities, rather than assuming a functional form for their influence on SNR.

\subsection{Matched-Filtering Singles and Multis} \label{sec:snr_match_examples}

\begin{figure*}[htb!]
    \centering
    \includegraphics[width=\textwidth]{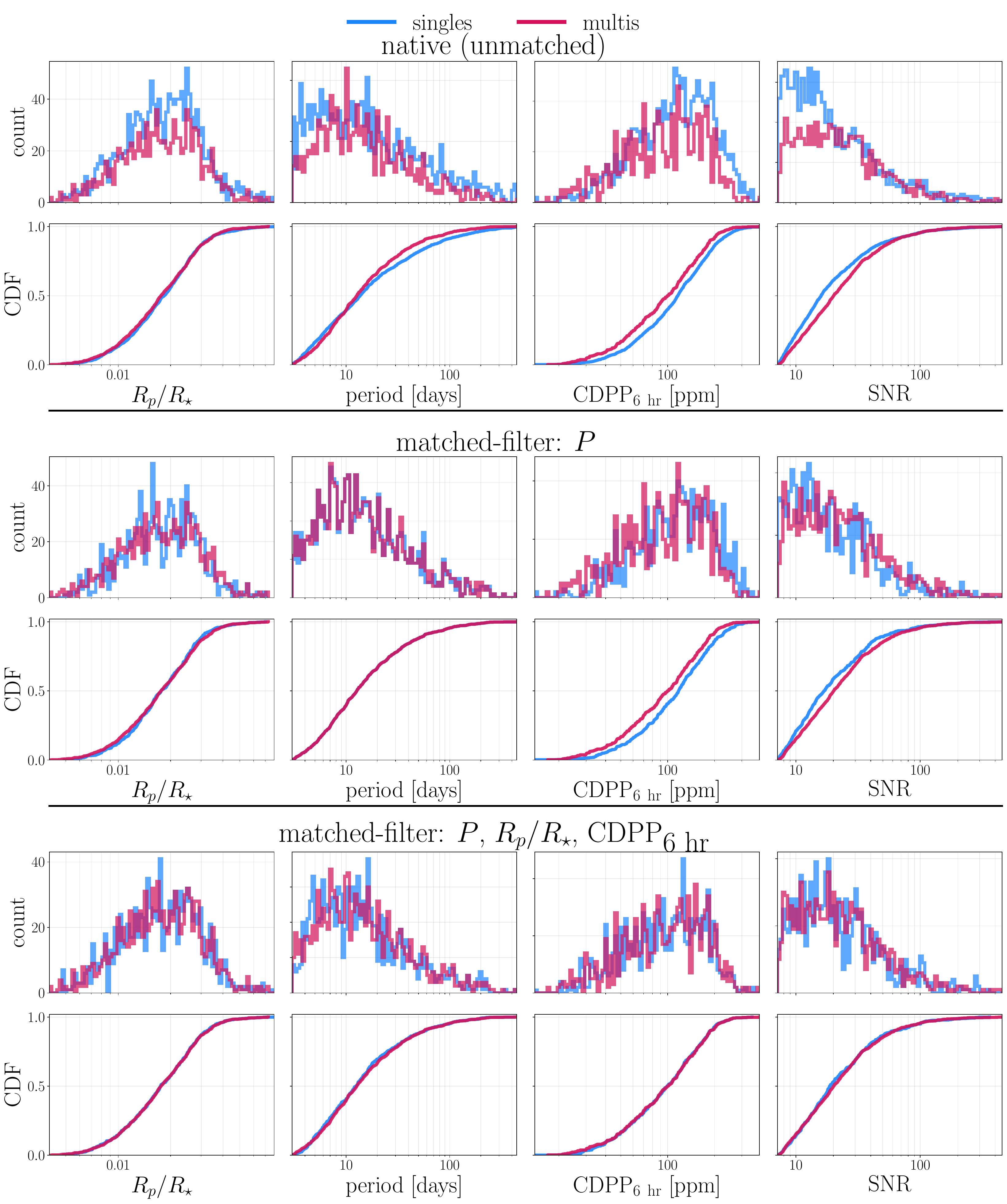}
    \caption{Distributions of $P$, $R_p/R_\star$, $\mathrm{CDPP}_{\rm 6\,hr}$, and SNR for single-planet (blue) and multi-planet (red) systems in $\mathcal{D}_{\rm kep}$ under three matching configurations. The top pair of rows shows the native (unmatched) population, the middle pair shows samples matched in orbital period ($P$), and the bottom pair shows samples matched jointly in $P$, $R_p/R_\star$, and $\mathrm{CDPP}_{\rm 6\,hr}$. Within each configuration, the upper row displays histograms of planets and the lower row shows the corresponding CDFs.}
    \label{fig:snr_match_examples}
\end{figure*}

To isolate differences in detectability between single- and multi-planet systems, we perform a suite of \emph{matched-filtered} comparisons in which the underlying distributions of key observables are explicitly controlled. 

If differences in the period distributions of singles and multis are a primary driver of the observed SNR offset, then matching the two populations in orbital period should lead to improved agreement between their SNR distributions. More generally, if the SNR differences are largely attributable to variation in period, $R_p/R_\star$, and stellar noise, then progressively matching in these parameters -- culminating in joint matching across all three -- should yield increasingly similar SNR distributions. 

Conversely, if period is not a dominant driver, then period-only matching is not expected to significantly alter the level of agreement. Likewise, if additional effects beyond these parameters contribute to the observed SNR differences (e.g., contamination or differing false positive rates), then discrepancies may persist even after matching in period, $R_p/R_\star$, and CDPP.

We implement a nearest-neighbor matched-filtering procedure in a multi-dimensional parameter space defined by combinations of orbital period ($P$), planet-to-star radius ratio ($R_p/R_\star$), and stellar noise, quantified by the 6-hour CDPP. For each choice of matching parameters, the relevant quantities are standardized before matching so that no single dimension dominates the Euclidean distance metric. Nearest-neighbor matching and feature standardization were implemented using the \texttt{scikit-learn} package \citep{scikit-learn}, specifically the \texttt{StandardScaler} and \texttt{NearestNeighbors} classes.

Multis are treated as the reference population. For each multi-planet candidate, we identify the nearest single-planet candidate in the selected parameter space using a Euclidean nearest-neighbor search. Matching is performed with replacement, so the same single-planet candidate may be selected as the nearest neighbor to more than one multi-planet candidate. The resulting matched-pair catalog therefore contains one nearest-neighbor single counterpart for each multi-planet candidate and a single-planet system can appear more than once if it is the nearest-neighbor of more than one multi-planet system.

We match over each of the three matching parameters ($P$, $R_p/R_\star$, and $\mathrm{CDPP}_{\rm 6\,hr}$) individually as well as the full three-parameter match. We also retain the native, unmatched singles and multis as a baseline comparison. 

Figure~\ref{fig:snr_match_examples} shows the $P$, $R_p/R_\star$, $\mathrm{CDPP}_{\rm 6\,hr}$, and SNR for single and multi-planet systems in $\mathcal{D}_{\rm kep}$ for three populations: the native (unmatched) population, period-only matched-filtering, and the full three-parameter matched-filtering. For each case, the top row shows histograms of individual systems and the bottom row shows the corresponding cumulative density functions (CDFs).

In the native population, the SNR distributions of singles and multis are visibly offset, as expected and similar to the result found in \citet{Lissauer2024} for all \textit{Kepler} planets. The main difference is that there is no longer an excess of high-SNR singles, because our $R_p < 7\,R_\oplus$ cut removes hot Jupiters. Matching in orbital period alone reduces this discrepancy, leading to clearly improved agreement between the two populations. When matching jointly in $P$, $R_p/R_\star$, and $\mathrm{CDPP}_{\rm 6\,hr}$, the SNR distributions become nearly identical, qualitatively.

For each matching configuration, we compare the SNR distributions of singles and multis using the two-sample Kolmogorov--Smirnov (KS) statistic,
\begin{equation}
D_{\rm KS} = \sup_x \left| F_{\rm singles}(x) - F_{\rm multis}(x) \right|,
\end{equation}
where $F_{\rm singles}$ and $F_{\rm multis}$ are the empirical CDFs of the two populations and a smaller $D_{\rm KS}$ indicates more similar distributions. We compute $D_{\rm KS}$ using \texttt{scipy.stats.ks\_2samp} \citep{2020SciPy-NMeth}. We use the KS statistic purely as a descriptive measure of the distance between the two SNR distributions and track how that distance changes as we match on different combinations of observables, rather than interpreting it as a formal hypothesis test.

\begin{figure}[htb!]
    \centering
    \includegraphics[width=\columnwidth]{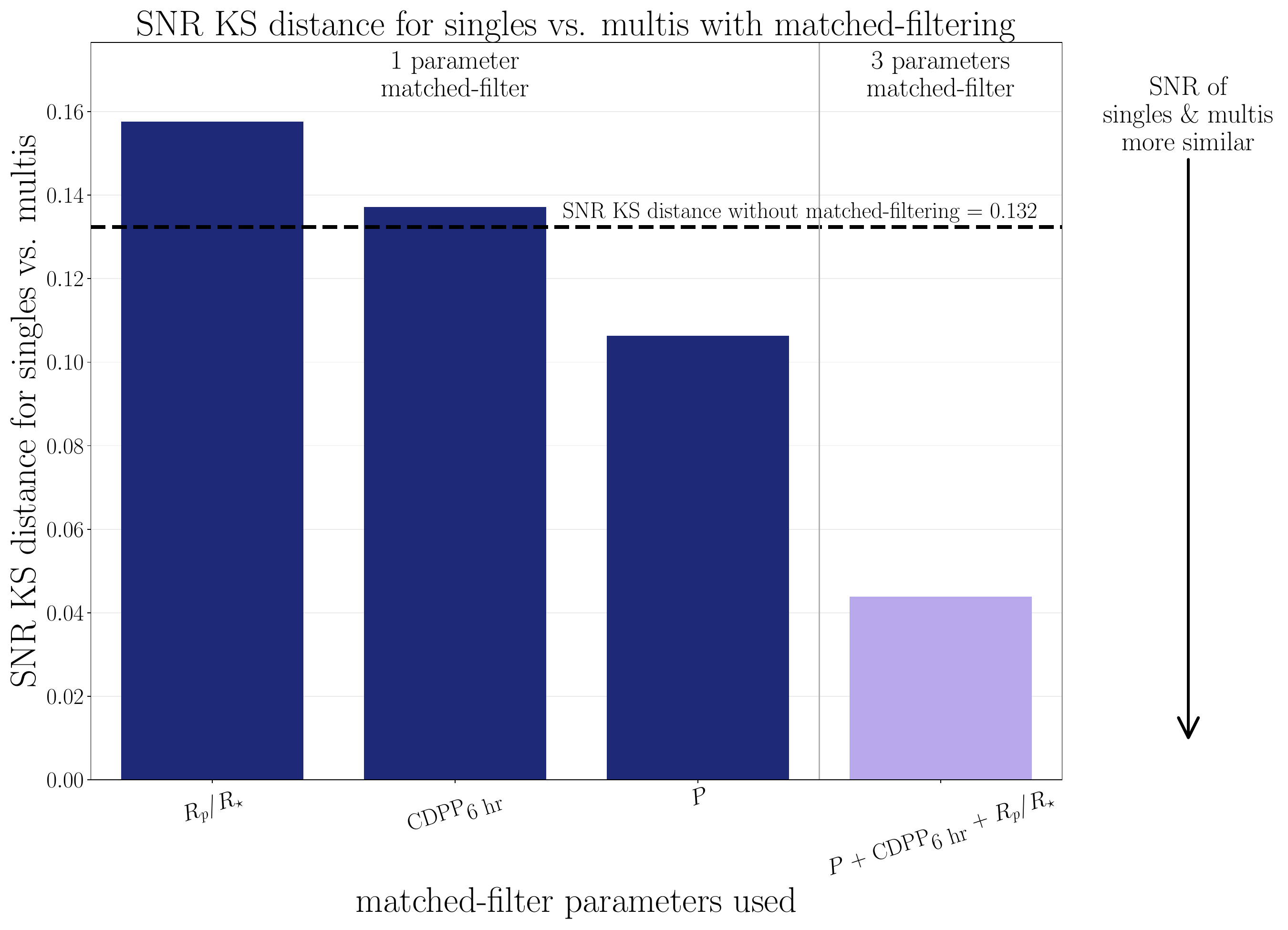}
    \caption{Kolmogorov-Smirnov (KS) statistics comparing the SNR distributions of singles and multis under different matched-filtering configurations. Dashed horizontal line shows the native KS distance of the SNR distributions without matched-filtering. For single parameter matched-filtering, only matching on period improves the agreement between SNR of singles and multis, while a matched-filter over all three parameters results in the smallest KS distance.}
    \label{fig:ks_comparison}
\end{figure}

Figure~\ref{fig:ks_comparison} shows $D_{\rm KS}$ between the single and multi SNR distributions for every matching configuration, along with the unmatched  baseline ($D_{\rm KS} = 0.132$). Of the three single-parameter matches, only matching in orbital period reduces the KS distance, to $D_{\rm KS} = 0.106$;  matching in $R_p/R_\star$ or $\mathrm{CDPP}_{\rm 6\,hr}$ alone leaves the two SNR distributions no more similar than the unmatched baseline, in fact slightly increasing the KS distance ($D_{\rm KS} = 0.157$ and $0.137$, respectively). Orbital period is therefore the only individual observable whose distribution differences, once removed, bring singles and multis into closer SNR agreement. 

The full three-parameter matched filter yields the smallest KS distance of any configuration, $D_{\rm KS} = 0.044$ -- roughly a third of the unmatched value -- rendering the two distributions nearly indistinguishable. Controlling jointly for period, radius ratio, and stellar photometric noise thus brings the singles' and multis' SNR distributions into substantially closer agreement, with orbital period the single most important of the three.

\section{Can Observational Selection Explain the Singles vs. Multis Difference?} \label{sec: intrinsic}

We have established that the SNR difference between observed single- and multi-tranets in \textit{Kepler} is driven primarily by differences in the orbital periods of their detected planets.

The \texttt{ExoMOD} forward model provides, for every simulated system, both its intrinsic architecture and the subset of planets that \textit{Kepler} would detect, so we can ask directly whether the observed single- vs. multi-tranet period difference is inherited from the underlying population or induced by the detection process. We restrict this analysis to systems in which at least one planet is detected -- the only systems that enter an observed catalog -- so that the intrinsic distributions we compare are those of \textit{detectable} systems, matched to what \textit{Kepler} actually samples. 

One caveat applies throughout the \texttt{ExoMOD} comparisons that follow. \texttt{ExoMOD} is calibrated to \textit{Kepler} planets with $P_{\rm orb} < 300$~d \citep{ExoMOD1}, whereas our \textit{Kepler} catalog extends to $500$~d (Section~\ref{sec: catalog_building}); the model therefore cannot populate the $300$--$500$~d tail, and the fractions we quote for the long-period single excess are evaluated within the model's $<\!300$~d domain. Importantly, this truncation suppresses precisely the long-period single-tranets that constitute the excess. Therefore it is conservative: it can only weaken, never manufacture, the long-period single excess we attribute to observational selection.

\begin{figure*}[t]
    \centering
    \includegraphics[width=\textwidth]{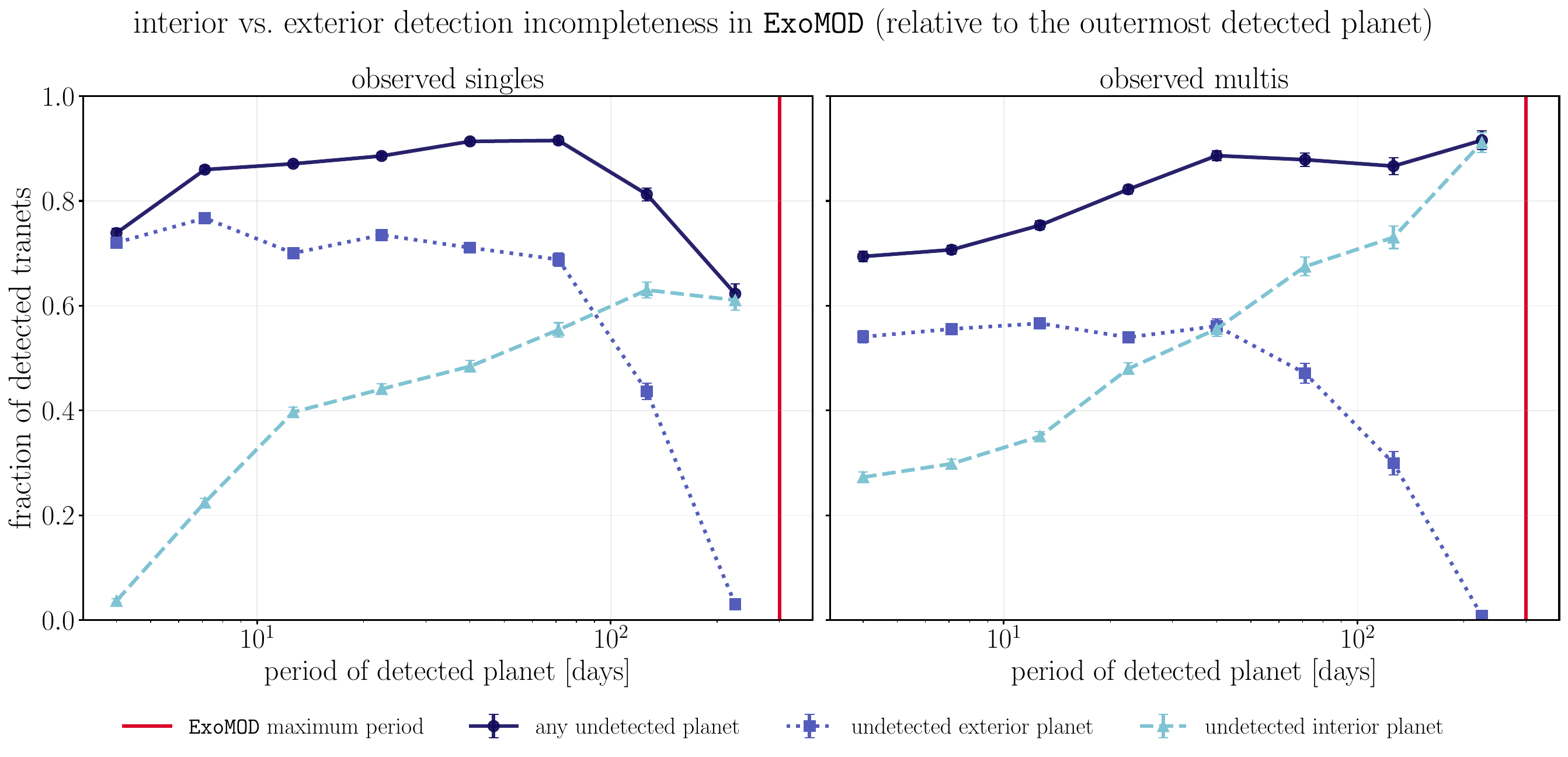}
    \caption{\texttt{ExoMOD} detection incompleteness as a function of the detected planet's orbital period, for systems cataloged as singles (left) and as multis (right). Solid curves show the fraction of detected tranets in systems with at least one undetected planet; for singles, this is the fraction that belong to intrinsically multi-planet systems (85\% overall). Dotted and dashed curves show the fraction with an undetected planet exterior and interior to the outermost detected planet, respectively. Exterior companions are artificially suppressed at long periods because \texttt{ExoMOD} contains no planets beyond 300\,d (red line). Error bars are binomial.}
    \label{fig:internal_external}
\end{figure*}

\citet{ExoMOD1} found that the majority (60--85\%, and specifically $85\%$ in $\mathcal{M}_{228}$) of detected single-tranets reside in intrinsically multi-planet systems, and that for a detected system there is a $\sim\!50\%$ chance \textit{Kepler} missed at least one inner or intermediate-period planet. Further, the detected \texttt{ExoMOD} populations reproduce the \textit{Kepler} DR25 period trend, with $10.4\%$ of detected singles beyond $100$ d compared to $5.1\%$ of detected multis -- a factor-of-two relative excess. This separation is not built into the underlying model as \texttt{ExoMOD} draws the innermost orbital period from the same distribution irrespective of system multiplicity \citep{ExoMOD1}, so it imposes no period law that distinguishes single- from multi-planet systems. The observed separation therefore emerges predominantly in the mapping from the intrinsic to the detected population -- i.e., from observational selection -- rather than being inherited from it.

Figure~\ref{fig:internal_external} resolves the detection incompleteness of \texttt{ExoMOD} systems split into observed singles and observed multis as a function of the detected planet's orbital period. Three features stand out.
 
First, detected singles are predominantly multis at every period. The fraction of single-tranets that belong to intrinsically multi-planet systems (solid curve, left panel) is $85\%$ averaged over all periods. Even among the long-period single-tranets that dominate the low-SNR excess ($P>100$\,d), $73\%$ arise from intrinsically multi-planet systems, so the excess itself is populated predominantly by multis whose companions escaped detection. The decline at the longest periods is likely caused by the truncation of the \texttt{ExoMOD} injected population at 300\,d.
 
Second, where the hidden companions lie depends on the period of the detected planet. Out to $\sim$100\,d, roughly 70\% of singles have an undetected companion exterior to the detected planet (dotted curve), while the fraction with an undetected interior companion (dashed curve) rises from $\sim$4\% for the shortest-period singles to $\sim$60\% beyond 100\,d, where interior companions dominate. (The steep drop in exterior companions beyond 100\,d is largely an artifact of the 300\,d truncation.) A short-period single is therefore typically the innermost member of a system whose outer planets went undetected, whereas a long-period single is often the most detectable member of a system whose interior companions were missed despite their shorter periods, most plausibly because they are smaller or do not transit. Both patterns follow from the order in which \textit{Kepler} detects planets, which we discuss in Section~\ref{sec: discussion}. 
 
Third, systems cataloged as multis are incomplete too. Relative to their outermost detected planet, $\sim$70--90\% have at least one undetected planet, with the same progression from exterior- to interior-dominated incompleteness (right panel). The main difference is that singles are more likely to hide an exterior companion, consistent with their lone detected planet usually being the innermost member of its system. Observed multiplicity therefore underestimates intrinsic multiplicity for most systems in both classes: in the model, singles and multis differ mainly in how many members of their systems \textit{Kepler} detected.

Taken together, these results indicate that the observed difference between the period distributions of single- and multi-tranets -- and specifically the excess of long-period, low-SNR single-tranets -- can be reproduced by observational selection effects, without invoking any intrinsic difference between the underlying populations of single- and multi-planet systems.\footnote{We note that an independent line of evidence, from a companion paper, is consistent with this picture. In \citet{ExoMOD2}, the population statistics of long-period TTVs ($P_{\rm TTV}/P_{\rm orb} > 10$) indicate that single transiting planets host dynamically interacting companions nearly as often as doubles, which would not be expected if observed singles were predominantly genuine single-planet systems.}

\section{Discussion} \label{sec: discussion}

\begin{figure*}[t]
    \centering
    \includegraphics[width=\textwidth]{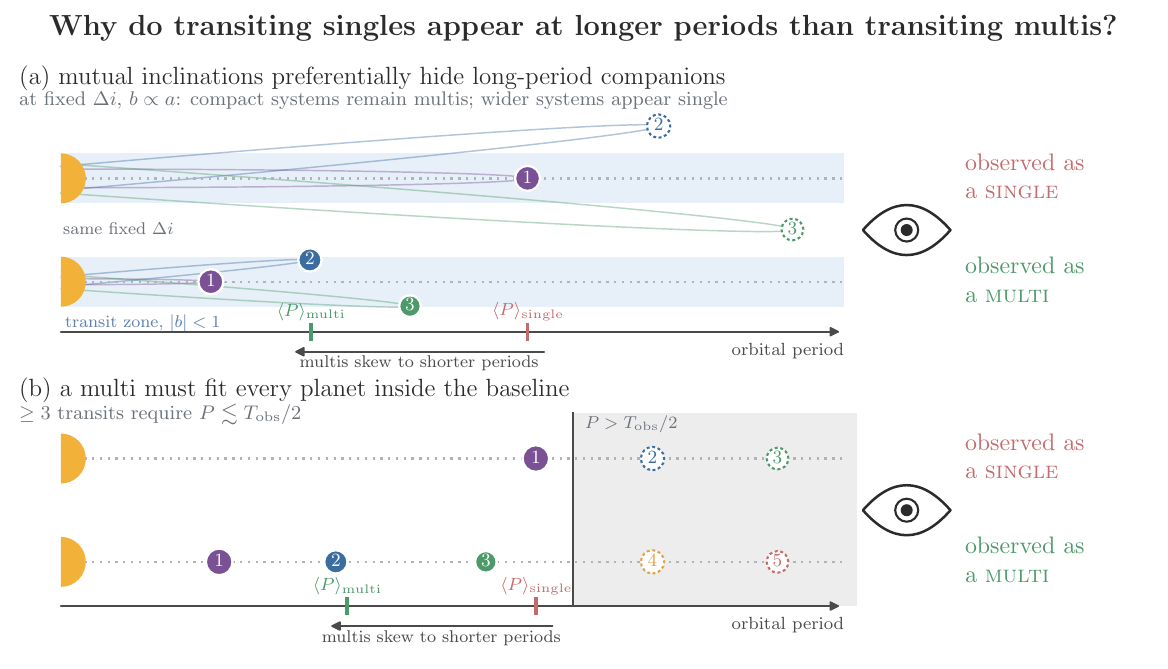}
    \caption{Two detection biases drive single-tranets from the same intrinsic planetary population to longer periods than multis: [a] mutual inclination lifts outer companions out of transit and [b] the 3-transit limit ($P \lesssim T_{\rm obs}/2$) packs multis to shorter $\langle P \rangle$.}
    \label{fig:baseline_bias_schematic}
\end{figure*}

A useful way to interpret these results is to ask, at a given orbital period, which planets \textit{Kepler} was capable of detecting at all. A candidate enters the DR25 catalog only if it satisfies three requirements, each of which depends on orbital period. First, the planet must transit: for a randomly oriented orbit the geometric transit probability scales as $p_{\rm tr} \approx R_\star/a \propto P^{-2/3}$, so transiting geometries become progressively rarer at longer periods. Second, the pipeline requires at least three transits within the observing baseline, imposing a hard upper limit $P \lesssim T_{\rm obs}/2$. Third, the folded signal must exceed the detection threshold (${\rm MES}\geq 7.1$). Because ${\rm SNR}\propto (R_p/R_\star)^2\,\sigma_0^{-1}\,P^{-1/3}$ (Equation~\ref{eq:snr_final}), this last condition defines a \emph{minimum detectable planet size} at each period. Setting the SNR equal to the detection threshold gives
\begin{equation} \label{eq:rmin}
    R_{\rm min}(P) \;\propto\; \sigma_0^{1/2}\,P^{1/6},
\end{equation}
so the smallest detectable planet grows -- slowly but monotonically -- with orbital period, and rises further around photometrically noisier stars. This behavior is borne out by the raw \textit{Kepler} candidate counts: a sharp edge is visible along the small-radius, long-period envelope of the observed radius--period distribution \citep[e.g.,][Fig.~1]{Latham2011}, arising because the number of in-transit cadences declines as $P^{-2/3}$.

Taken together, these three conditions carve out the region of the period--radius plane that \textit{Kepler} could sample. At short periods a broad range of planet sizes is both likely to transit and comfortably above threshold; at longer periods only larger planets, in the diminishing fraction of systems oriented edge-on enough to transit, clear the detection floor before the three-transit baseline truncates the sample entirely. The detectable population is therefore increasingly restricted to large, short-period planets, and the first planet recovered in any system is preferentially the one that sits furthest above this boundary -- that is, the largest and/or shortest-period member. Additional planets emerge only as more data accumulate or as one looks toward smaller sizes and longer periods, where $R_{\rm min}(P)$, the transit probability, and the baseline limit all work against detection.

Our own data bear out this ordering directly. We construct a subset of $\mathcal{D}_{\rm kep}$ consisting of systems present in the \textit{Kepler} half-mission (quarters 1--8) catalog by downloading that catalog from the NASA Exoplanet Archive and cross-matching on \textit{Kepler} ID. Among systems classified as single in the first half of the mission but later found to host multiple planets in the full DR25 catalog, every newly detected planet with a radius comparable to or larger than the originally known planet lies at a \emph{longer} orbital period than that planet. Although this subset is modest ($17$ systems), the complete absence of counterexamples indicates that equal-size or larger companions are preferentially found exterior to the initially detected planet -- exactly as expected if detection proceeds inward from the most favorable corner of the period--radius plane. 

A complementary view comes from the relative sizes of planets uncovered by extending the \textit{Kepler} baseline: of the planets found interior to the originally known planet, $47/47$ ($100\%$) are smaller than that planet, whereas of those found exterior, only $29/46$ ($63\%$) are smaller. This asymmetry reinforces the same detection-order bias: the first planet recovered in a system is typically the largest and highest-SNR signal. 

This framework provides a natural explanation for why single-transiting systems are observed at longer periods, on average, than multis -- without invoking any intrinsic difference between the underlying single- and multi-planet populations. The key point is that being \emph{cataloged as a multi} is a far more restrictive condition than being cataloged as a single, and that restriction acts most strongly at long periods. Two effects, each illustrated in Figure~\ref{fig:baseline_bias_schematic}, contribute.

First, transit geometry (see panel [a] of Figure~\ref{fig:baseline_bias_schematic}). In a multi-planet system, non-zero mutual inclinations mean that not all planets transit; because the impact parameter grows as $b \simeq (a/R_\star)\sin\Delta i$, a given mutual inclination lifts the outer, longer-period companions out of transit first. A modest tilt therefore leaves only the innermost planet transiting, and this effect strengthens with period, so long-period detections are preferentially seen as singles. Second, the finite baseline (see panel [b] of Figure~\ref{fig:baseline_bias_schematic}). The longest recoverable period satisfies $P\lesssim T_{\rm obs}/2$. A single-planet system may place its lone detected planet anywhere up to this limit, but a multi must fit \emph{every} detected planet, each with three transits, inside the same baseline. The average detected period of a multi is therefore mechanically pushed to shorter values, and increasingly so at higher multiplicity.

The \texttt{ExoMOD} population lets us verify that the transit geometry and mutual inclinations follow these expectations. We contrast intrinsically multi-planet systems cataloged as singles with those cataloged as multis: the former are more mutually inclined, with a median line-of-sight inclination dispersion of $1.97^\circ$ versus $1.05^\circ$. Consistent with $b\simeq(a/R_\star)\sin\Delta i$, $83\%$ of the undetected companions of cataloged singles are non-transiting and the majority lie exterior to the detected planet, confirming that inclination preferentially lifts the outer companions out of transit.

Additionally, if singles are merely multis whose innermost detected planet happens to fall at longer intrinsic period, the two classes should carry a comparable rate of exterior, transiting-but-undetected companions -- and they do, with nearly the same fraction hosting such a companion ($11.1\%$ of singles versus $12.2\%$ of multis). Because \texttt{ExoMOD} is built on a single underlying population, this equality is a consistency check confirming that the exterior-companion channel behaves as the detection-order picture predicts. The two classes differ mainly in where along the period axis their innermost detected planet falls, not in their access to hidden exterior companions -- as expected if cataloged singles are drawn from the same architectures as multis. 

Mechanism (b) suggests a further test that lies beyond the scope of the present work. Multiplicity greater than one is awarded only to systems with two or more planets inside the baseline-limited detection window, so the companions (b) hides are not non-transiting but under-sampled. Thus, they should sometimes appear as single- or double-transit events. Cataloged singles should therefore host mono- and duo-transiting companions -- exterior to the detected planet -- at a rate set by the intrinsic multiplicity function. The prediction is an absolute rate, not an excess over multis, which are equally free to hold planets beyond their own outermost detection. Measuring it would help separate a genuine architectural edge \citep[e.g.,][]{Millholland2022} from one imposed by the requirement of repeat transits.

We emphasize that these results establish sufficiency rather than necessity: they show that observational selection acting on a single parent population is enough to reproduce the observed single--multi differences, not that no intrinsic difference between the two populations is possible. Observational selection acting on a single parent population is enough to reproduce the longer periods, and hence the lower SNRs, of cataloged singles -- so the low-SNR excess is not itself evidence for a \textit{Kepler} dichotomy. These results cannot, however, exclude an intrinsic architectural difference between singles and multis, such as that predicted by formation models in which the extent of migration sets how many planets a system leaves inside the detection window \citep[e.g.,][]{Zawadzki2022}.

\section{Conclusions} \label{sec: conclusions}

Motivated by the excess of low-SNR single-transiting systems reported by \citet{Lissauer2024}, we investigate whether this excess originates in the false positive rates, the detection (true positive) rates, or the intrinsic populations of \textit{Kepler} singles and multis. Analyzing a DR25 FGK-dwarf catalog of $2{,}557$ planet candidates and the false positive-free \texttt{ExoMOD} forward model, we reach the following conclusions.

\begin{enumerate}
    \item The fraction of low-SNR detections declines monotonically with observed multiplicity -- from singles through doubles, triples, and $4+$ systems -- and this same ordered trend is reproduced by the \texttt{ExoMOD} detected population, which contains no false positives. The low-SNR excess in singles therefore does not require false positive contamination, though the larger drop from singles to doubles in the real data leaves room for a modest additional contribution.

    \item The SNR offset between singles and multis is driven predominantly by differences in their orbital-period distributions. Among single-parameter matches, only matching in orbital period brings the two SNR distributions closer together, whereas matching in $R_p/R_\star$ or $\mathrm{CDPP}_{\rm 6\,hr}$ alone does not; and the full match in all three observables simultaneously brings the two SNR distributions to near-agreement. An independent Monte Carlo resampling, in which the singles are given the multis' orbital-period distribution while their non-period SNR normalization is held fixed, reproduces the observed multi SNR distribution -- corroborating, without any nearest-neighbor matching, that period differences account for much of the offset.

    \item The observed excess of long-period single-tranets is reproduced by \texttt{ExoMOD} without any multiplicity-dependent period law. In the model, $85\%$ of detected single-tranets -- and $73\%$ of those beyond $100$~d -- belong to intrinsically multi-planet systems in which \textit{Kepler} detected only one member. Thus, the observed period difference between single- and multi-tranet systems is explainable from detection effects rather than from an intrinsic difference between the two populations.

    \item These biases follow directly from the period dependence of detectability. At fixed stellar noise the minimum detectable planet size grows as $R_{\rm min}\propto P^{1/6}$, transit probability falls as $P^{-2/3}$, and the three-transit requirement caps detections at $P\lesssim T_{\rm obs}/2$. Together these push the detected multi population toward short periods and channel long-period, low-SNR detections into the observed-single category.
\end{enumerate}

Taken together, these results indicate that the low-SNR excess of \textit{Kepler} singles -- and the associated excess of long-period singles -- does not require false positives and can arise from observational selection resulting in longer-period singles, rather than from an intrinsically distinct population of single-planet systems. In this sense, much of the apparent differences between the SNRs and periods of \textit{Kepler} singles and multis is a manifestation of how transiting planets are detected rather than of how planetary systems are built.

\begin{acknowledgements}
We are grateful to Eric Agol, Megan Bedell, Andrew Casey, Jessie Christiansen, Julien de Wit, Gregory Gilbert, David Hogg, Tiger Lu, Adrian Price-Whelan, Malena Rice, Avi Shporer, and Quang Tran for fruitful discussions. This work benefited from insightful conversations with Astronomical Data Group at the Flatiron Institute Center for Computational Astrophysics, the Millholland Research Group, and the Vanderburg Research Group. D.A.Y. is supported by a Juan Carlos Torres Postdoctoral Fellowship at the Massachusetts Institute of Technology.

This research has made use of the NASA Exoplanet Archive, which is operated by the California Institute of Technology, under contract with the National Aeronautics and Space Administration under the Exoplanet Exploration Program.

This work made use of the following software packages: \texttt{Jupyter} \citep{2007CSE.....9c..21P,kluyver2016jupyter}, \texttt{matplotlib} \citep{Hunter:2007}, \texttt{numpy} \citep{numpy}, \texttt{pandas} \citep{mckinney-proc-scipy-2010,pandas_17229934}, \texttt{python} \citep{python}, and \texttt{scipy} \citep{2020SciPy-NMeth,scipy_17101542}.

Software citation information aggregated using \texttt{\href{https://www.tomwagg.com/software-citation-station/}{The Software Citation Station}} \citep{software-citation-station-paper,software-citation-station-zenodo}.

During the preparation of this work, the authors used OpenAI's ChatGPT and Codex and Anthropic's Claude models to assist with code development and writing improvements. All AI-assisted output was reviewed, edited, and verified by the authors, who take full responsibility for the content of this manuscript.

\end{acknowledgements}


\bibliography{main}{}
\bibliographystyle{aasjournalv7}


\end{CJK*}
\end{document}